\documentstyle[11pt,moriond,epsfig]{article}

\def\Journal#1#2#3#4{{#1} {\bf #2}, #3 (#4)}

\def\NPB{{\em Nucl. Phys.} B}
\def\PLB{{\em Phys. Lett.}  B}

\def\PRD{{\em Phys. Rev.} D}

\def\be{\begin{equation}}
\def\ee{\end{equation}}
\def\bea{\begin{eqnarray}}
\def\eea{\end{eqnarray}}

\begin{document}
\vspace*{4cm}
\title{MONOPOLES VORTICES AND CONFINEMENT}

\author{ A. DI GIACOMO }

\address{Dipartimento di Fisica \& I.N.F.N. Pisa, Via Buonarroti 2,
\\ 56100 Pisa, Italia}

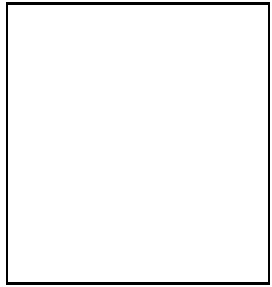
\maketitle\abstracts{
The status of our understanding of colour confinement is reviewed.}

\section{Introduction}
Confinement of colour is a fundamental problem in quantum field 
theory. Understanding the mechanism
of confinement can also suggest observable predictions for heavy ion 
collisions.

Phenomenology strongly indicates that quarks exist as fundamental 
particles of hadronic
matter. However quarks have never been observed as free particles.
Upper limits to their production cross section, $\sigma_q$, are reported
by the particle
data group\cite{1}.
For reactions initiated by protons
$\sigma_q \leq 10^{-40}\,{\rm cm}^2$.
The ratio to the total cross section $\sigma_{Tot} \simeq 
10^{-25}\,{\rm cm}^2$ has then the bound
${\sigma_q}/{\sigma_{Tot}}\leq 10^{-15}$.
Similarly the expected abundance of relic quarks in the standard cosmological
model as compared to
the abundance of nucleons is expected to be\cite{2}
${n_q}/{n_p} \geq 10^{-12}$.
The experimental upper limit produced by Millikan like experiments looking for
fractionally
charged particles is
${n_q}/{n_p} \leq 10^{-27}$,
corresponding to zero quarks observed in $\sim 1$~g of matter. Again
$n_{q,obs}/n_{q,exp}\leq 10^{-15}$.
A non zero value of the above ratios, smaller than $10^{-15}$ would have no
natural theoretical
explanation. The most natural possibility is that they are equal to zero,
and that confinement is
an absolute property to be explained in terms of symmetry\cite{3}.

Perturbative quantization is not adequate to describe long distance 
physics: a non perturbative
formulation like lattice can instead provide answers from first principles.
Lattice does indeed indicate that colour is confined below a 
transition temperature $T\sim
150$~MeV, at which a deconfining transition takes place to a phase of 
quark gluon plasma\cite{}.
The parameters used to detect confinement are either the string 
tension $\sigma$, which enters in
the long range part of the $q\bar q$ potential:
$ V(R)\mathop\sim_{R\to\infty} \sigma R $,
and is extracted from the numerical determinations of Wilson loops; 
or the expectation value
$\langle L\rangle$ of the Polyakov line $L(x)$ (the parallel 
transport from $t=-\infty$ to
$t=\+\infty$ along the time axis. $\langle L\rangle \sim 
\exp(-\zeta/T)$, $\zeta$ being the
chemical potential of a quark in the vacuum: $\langle L\rangle = 0$ 
means $\zeta = \infty$, i.e.
confinement. Lattice simulations at finite temperature are made on a 
lattice $N_T\times N_S^3$
($N_S\gg N_T$), the temperature $T$ being related to the lattice spacing as
$T = {1}/{a N_T}$.
Since
\begin{equation}
a\mathop\simeq_{\beta\to\infty} \frac{1}{\Lambda_L}\exp(-b_0\beta)
\end{equation}
with $\beta = 2 N_c/g^2$ and $b_0 > 0$ (asymptotic freedom), low temperature
(confined phase)
corresponds to strong
coupling. Explaining confinement in terms of symmetry, means then 
studying the symmetry of the
strong coupling (disordered) phase. The key word for that is duality\cite{4}.
\section{Duality.}
Duality is a deep concept in statistical mechanics and in quantum 
field theory. It applies to
systems admitting non local excitations with non trivial topology. 
Such systems admit two
complementary descriptions. A {\em direct} description in terms of 
fields $\Phi$, which is
suitable in the weak coupling regime,
$g
\ll 1$ (ordered phase). Their v.e.v. $\langle\Phi\rangle$ are called
order parameters: in this description there exist non local 
excitations $\mu$. A {\em dual}
description, suitable in the strong coupling regime (disordered 
phase), in which the
excitations $\mu$ become local, and the fields $\Phi$ non local. The 
symmetry is described in
terms of $\langle\mu\rangle$, the disordered parameters.
The ``dual'' effective coupling $g_D$ is related to the direct $g$ by 
the relation $g_D\sim 1/g$:
duality maps the strong coupling regime of one description into the 
weak coupling regime of the
other.

Examples of system with duality are:
\begin{itemize}
\item[1)] The 2d Ising model, where $\Phi$ is the spin, $\mu$ are the 
kinks\cite{5,6}.
\item[2)] The $N=2$ SUSY QCD, where $\Phi$ are the gauge superfields, 
$\mu$ monopoles\cite{7}.
\item[3)] The 3d Heisenberg magnet, where $\Phi$ is the 
magnetization, $\mu$ are the Weiss
domains\cite{8}.
\item[4)] The 3d $XY$ model (liquid $He_4$) where $\Phi$ are the 
velocities, $\mu$ the
vortices\cite{9}.
\item[5)] Compact $U(1)$ gauge theory, where $\Phi$ is the e.m. 
field, $\mu$ are monopoles
\cite{10}.
\end{itemize}
The problem for $QCD$ is to find the dual excitations $\mu$, which 
condense in the confined phase,
and are weakly interacting.
A less ambitious task is to understand the symmetry of the confining 
vacuum. The keywords in this
game are vortices and monopoles.

\section{Monopoles}
The natural topology in 3 dimensions comes from a mapping of the 
sphere $S_2$ at infinity on a
group. A mapping of $S_2$ on $SO(3)/U(1)$, or on $SU(N)/SU(p)\otimes 
SU(N-p)\otimes U(1)$,
$(p=1\ldots N-1)$ has monopoles as topological excitations. If the 
dual excitations $\mu$ have
nonzero magnetic charge, $\langle\mu\rangle\neq 0$ signals dual 
superconductivity. Then
confinement is produced by the squeezing of the chromoelectric field 
of a $q\bar q$ pair into
Abrikosov like flux tubes. The energy is proportional to the length, so that
$V = \sigma R$.
A conserved magnetic charge can be defined in a gauge theory by a 
procedure known as abelian
projection\cite{11}.
Choosing $SU(2)$ for simplicity of notation, consider a field $\vec\Phi(x)$
belonging to the adjoint representation, i.e. a vector in colour 
space. Define $\hat
\Phi(x)\equiv \vec\Phi(x)/|\vec\Phi(x)|$: this can be done everywhere 
except at zeros of
$\vec\Phi(x)$.
Define a field strength tensor
\begin{equation}
F_{\mu\nu} = \hat\Phi\cdot\vec G_{\mu\nu} - 
\frac{1}{g}\hat\Phi\cdot(D_\mu\hat\Phi\wedge
D_\nu\hat\Phi)\label{hfield}
\end{equation}
Here $\vec G_{\mu\nu} = \partial_\mu\vec A_\nu - \partial_\nu\vec A_\mu + g
\vec A_\mu\wedge\vec A_\nu$, $D_\mu = \partial_\mu + g\vec 
A_\mu\wedge$ is the covariant
derivative.

$F_{\mu\nu}$ is a colour singlet and gauge invariant. In fact both 
terms in eq.(\ref{hfield}) are
separately gauge invariant: the choice of the coefficients is such 
that bilinear terms in $A_\mu
A_\nu$  cancel. By simple algebra
\begin{equation}
F_{\mu\nu} = \hat\Phi\cdot(\partial_\mu\vec A_\nu - \partial_\nu\vec 
A_\mu) - \frac{1}{g}
\hat\Phi(\partial_\mu\hat\Phi\wedge \partial_\nu\hat\Phi)\label{hfield2}
\end{equation}
$\hat\Phi(x)$ can be made a constant vector by a gauge 
transformation: the second term in
eq.(\ref{hfield2}) then drops and $F_{\mu\nu}$ becomes an abelian field
\begin{equation}
F_{\mu\nu} = \partial_\mu(\hat\Phi\vec A_\nu) - 
\partial_\nu(\hat\Phi\vec A_\mu)
\end{equation}
Such a gauge transformation
is called abelian projection of $\hat\Phi$: it is singular at the 
zeros of $\vec\Phi(x)$, where
$\hat\Phi$ is not defined. The dual tensor
$F^*_{\mu\nu} = \frac{1}{2}\varepsilon_{\mu\nu\rho\sigma} F^{\rho\sigma}$
defines a magnetic current
$j_\mu = \partial^\nu F^*_{\nu\mu}$
which is identically conserved, being $F^*_{\mu\nu}$ antisymmetric. 
$j_\mu$ is usually zero
(Bianchi identities): it can be non zero in the compact formulation 
of the theory.
Monopoles sit then at the zeros of $\vec\Phi(x)$.

There exists a conserved magnetic charge for any choice of the field 
$\vec\Phi(x)$.

If an operator $\mu$ carrying nonzero magnetic charge has non zero 
v.e.v., $\langle\mu\rangle\neq
0$, the $U(1)$ magnetic simmetry is broken \`a la Higgs, and this 
implies dual superconductivity.

A magnetically charged operator $\mu$ can be constructed\cite{12}. 
The construction is the
adaptation to a compact theory of the translation operator $e^{ipx}$
\begin{equation}
e^{ipa}|x\rangle = |x + a\rangle
\end{equation}
In the gauge theory $x$ is the field, $p$ the conjugate momentum, $a$ 
a classical monopole
configuration: the translation operator adds a classical monopole to any field
configuration\cite{12,13}.

$\langle\mu\rangle$ is then determined by numerical simulations as a 
function of the temperature.

If the idea of dual superconductivity as a mechanism for confinement 
is correct,
$\langle\mu\rangle$ should be different from zero in the confined 
phase, and go to zero in the
deconfined phase. Of course this is strictly true in the infinite 
volume limit, when a phase
transition can exist.

This is exactly what happens: $\langle\mu\rangle$ is determined at 
different sizes of the
spatial size of the lattice, and the extrapolation to infinite volume 
is done by finite size
scaling analysis.

The result is that $\langle\mu\rangle\neq 0$ $T < T_c$, 
$\langle\mu\rangle = 0$ $T > T_c$.
Moreover $\langle\mu\rangle \simeq (1- T/T_c)^\delta$, $\delta = 0.5\pm.1$.

The result is independent of the abelian projection. Dual 
superconductivity is the mechanism
of colour confinement. The same analysis can be made in the presence 
of quarks: contrary to the
string tension, which is not defined in this case, because the string 
breaks producing $q\bar q$
pairs, or to $\langle L\rangle$ which also is not defined in this 
case because quarks break the
$Z_N$ symmetry, dual superconductivity is  a property of the vacuum 
which can exist.
The study of $\langle\mu\rangle$ can shed light on the relation 
between chiral phase transition
and deconfinement transition.
\section{Vortices.} Vortices in $3+1$ dimensions are defects 
associated to closed lines, $C$,
which are a kind of ``dual Wilson loops''. They obey the relation 
with Wilson loop $W(C')$
\begin{equation}
B(C) W(C') = W(C') B(C) \exp(i\frac{n_{CC'}}{N}2\pi)\label{comm}
\end{equation}
In the same way as $\langle W(C')\rangle \neq 0$, $\langle 
B(C)\rangle\neq 0$ has no special
meaning.
$B(C)$ does not carry any conserved quantum number. What follows from 
eq.(\ref{comm}) is that if
$\langle B(C)\rangle$ obey the area law $\langle W(C)\rangle$ obey 
the perimeter law and viceversa.

This has been checked on the lattice\cite{14}.
\section{Concluding remarks.}
The QCD vacuum in the confined phase is a dual superconductor for the 
$U(1)$ magnetic symmetry
defined by any abelian projection. This result is for sure a step in 
the direction of understanding
confinement. It  says that the dual excitations have magnetic charge 
in all the abelian
projections. The details of these excitations are, however, still unknown.

\end{document}